\documentclass[aps,nofootinbib,twocolumn,prl,showpacs,class-pre]{revtex4}
\usepackage{setspace} 
\usepackage{amsmath,amssymb,amsfonts,amsthm}
\usepackage{graphicx}

\newcommand{\be}{\begin{equation}}
\newcommand{\ee}{\end{equation}}
\newcommand{\beq}{\begin{eqnarray}}
\newcommand{\eeq}{\end{eqnarray}}
\newcommand{\br}{{\bf r}}
\newcommand{\bR}{{\bf R}}

\def\lsim{\hbox{ \raise.35ex\rlap{$<$}\lower.6ex\hbox{$\sim$}\ }}
\def\gsim{\hbox{ \raise.35ex\rlap{$>$}\lower.6ex\hbox{$\sim$}\ }}

\begin{document}
\title{Constraining the Noncommutative Spectral Action via
  Astrophysical Observations} \author{William
  Nelson\footnote{nelson@gravity.psu.edu}, Joseph
  Ochoa\footnote{jro166@psu.edu}} \affiliation{ Institute of
  Gravitation and the Cosmos, Penn State University, State College, PA
  16801, U.S.A.}  \author{Mairi
  Sakellariadou\footnote{mairi.sakellariadou@kcl.ac.uk}}
\affiliation{Department of Physics, King's College, University of
  London, Strand WC2R 2LS, London, U.K.}

\begin{abstract}
The noncommutative spectral action extends our familiar notion of
commutative spaces, using the data encoded in a spectral triple on an
almost commutative space. Varying a rather simple action, one can
derive all of the standard model of particle physics in this setting,
in addition to a modified version of Einstein-Hilbert gravity.  
In this letter we use
observations of pulsar timings, assuming that no deviation from
General Relativity has been observed, to constrain the gravitational
sector of this theory.  Whilst the bounds on the coupling constants
remain rather weak, they are comparable to existing bounds on
deviations from General Relativity in other settings and are likely to
be further constrained by future observations.
\end{abstract}


\maketitle

\section{Introduction}
Approaching Planckian energies, the assumption of Riemannian geometry
and the validity of General Relativity (GR) break down and one is forced to
describe the space-time geometry within a fully quantum
framework. NonCommutative Geometry (NCG)~\cite{ncg-book1,ncg-book2} is
based on the idea that as we approach Planckian energy scales, our
intuitive description of space-time being a commutative manifold
ceases to be a valid approximation.  In its simplest but nevertheless
powerful version, NCG implies that slightly below Planck energy,
space-time is well approximated by the product of a four-dimensional
smooth compact Riemannian manifold ${\cal M}$ and a finite
noncommutative space ${\cal F}$. Such spaces are called ``almost
commutative'' spaces and they are the simplest extensions of the
commutative spaces we use in GR. This is a strong
assumption which is certainly expected to break at the Planck scale,
where the notion of classical geometry loses all meaning, however at
low energies it should be a good approximation.

The noncommutative nature
of ${\cal F}$ is given by the real spectral triple $({\cal A}, {\cal
  H}, D)$, where ${\cal A}$ is an involution of operators on the
Hilbert space ${\cal H}$, and $D$ is a self-adjoint unbounded
operator in ${\cal H}$. The algebra ${\cal A}$ is the algebra of
coordinates, the operator $D$ corresponds to the inverse line element
of Riemannian geometry, and the commutator $[D,a]$ with $a\in {\cal
  A}$ plays the r\^ole of the differential quotient ${\rm d}a/{\rm
  d}s$, with ${\rm d}s$ the unit of length. The resulting physical
Lagrangian is obtained from the asymptotic expansion in the energy
scale $\Lambda$ of a spectral action functional of the form
%
${\rm  Tr}(f\big(D/\Lambda))$
%
 defined on noncommutative spaces, where $f$ is a cut-off function (i.e.\ a
test function of compact support).  The
 coupling with fermions can be obtained by including an additional
 term in the spectral action functional.
The
 choice of the finite dimensional algebra is the underlying geometric
 input which determines the physical implications of the model, in 
particular the particle content of the theory. 

The NCG spectral action offers a simple and elegant explanation for
the phenomenology of the Standard Model (SM) compatible with
right-handed neutrinos and neutrino masses~\cite{ccm} (the approach
has also been used to derive supersymmetric extensions to the standard
model~\cite{Broek:2010jw}). This approach to the SM has been proposed
as a way to achieve unification, based on the symplectic-unitary group
(the algebra constructed in ${\cal M}\times {\cal F}$ is assumed to be
symplectic-unitary) in the Hilbert space, instead of the finite
dimensional Lie groups. Note that the NCG spectral action is a
classical theory which, in principle, should eventually be
quantized. Whilst an understanding of how to quantize such
noncommutative spaces has not yet been fully developed, already at the
classical level the theory introduces several extensions to standard
GR.  Specifically, the physical Lagrangian contains,
in addition to the full SM Lagrangian, the Einstein-Hilbert action
with a cosmological term, a topological term related to the Euler
characteristic of the space-time manifold, a conformal Weyl term and a
conformal coupling of the Higgs field to gravity. In
contrast to the SM on a fixed background, the coefficients of the
gravitational terms in this NCG action depend on the Yukawa parameters
of the particle physics content.

The parameters of the NCG spectral action model are set at the scale
$\Lambda$, considered to be the unification scale, while physical
predictions at lower energies are recovered by running the parameters
down through Renormalization Group Equations (RGE). Thus, the spectral
action at the unification scale $\Lambda$ is directly applicable to
early universe cosmological
models~\cite{Nelson:2008uy,Nelson:2009wr,Marcolli:2009in,mmm}, while
extrapolations to lower energies can be obtained via RGE and inclusion
of nonperturbative effects in the spectral action.

The gravitational part of the asymptotic formula for the bosonic
sector of the NCG spectral action, including the coupling between the
Higgs field and the Ricci curvature scalar, reads~\cite{ccm}
\beq\label{eq:0}{\cal S}_{\rm grav} = \int \left(
\frac{1}{2\kappa_0^2} R + \alpha_0
C_{\mu\nu\rho\sigma}C^{\mu\nu\rho\sigma} + \tau_0 R^\star
R^\star\right.  \nonumber\\ -\left.  \xi_0 R|{\bf H}|^2 \right)
\sqrt{-g} {\rm d}^4 x~; \eeq
${\bf H}$ is a rescaling ${\bf H}=(\sqrt{af_0}/\pi)\phi$ of the Higgs
field $\phi$ to normalize the kinetic energy, the momentum $f_0$ is
physically related to the coupling constants at unification and the
coefficient $a$ is related to the fermion and lepton masses and
lepton mixing. Note that 
we are using conventions in which the signature is
$(-,+,+,+)$ and the Ricci tensor is defined as $R_{\mu\nu} = 
R^\rho\phantom{}_{\mu\nu\rho}$,
with $R_{\mu\nu\rho}\phantom{}^\sigma\omega_\sigma =
 \big[ \bigtriangledown_\mu , \bigtriangledown_\nu \big] \omega_\rho$. 
In the above action, Eq.~(\ref{eq:0}), the first term is the familiar
Einstein-Hilbert term, the second one is a Weyl curvature term, the
third term $R^\star
R^\star=(1/4)\epsilon^{\mu\nu\rho\sigma}\epsilon_{\alpha\beta\gamma\delta}
R^{\alpha\beta}_{\mu\nu}R^{\gamma\delta}_{\rho\sigma}, $ is the
topological term that integrates to the Euler characteristic, hence
nondynamical, and the last one couples gravity to the Higgs
field\footnote{ Such a term should always be present when one
  considers gravity coupled to scalar fields.} and can have important
consequences at high energies, such as in the early
universe~\cite{Nelson:2008uy,Nelson:2009wr,Marcolli:2009in,mmm}.
Here, we will be concerned with the low energy, weak curvature regime
where this term is small.

Neglecting the nonminimal coupling between the Higgs field and the
Ricci curvature, the equations of motion derived from the
spectral action above read~\cite{Nelson:2008uy}
\beq\label{eq:EoM2} R^{\mu\nu} - \frac{1}{2}g^{\mu\nu} R +
 \frac{1}{\beta^2} \left[ 2C^{\mu\lambda\nu\kappa}_{;\lambda ; \kappa}
  + C^{\mu\lambda\nu\kappa}R_{\lambda \kappa}\right]\nonumber\\ =   
\ 8\pi G T^{\mu\nu}_{\rm matter}~,  \eeq                            
where $\beta^2$ is defined as $\beta^2 = -1/(32\pi G
\alpha_0)$. Notice in particular that the NCG corrections vanish for
Friedmann-Lema\^{i}tre-Robertson-Walker (FLRW)
cosmologies~\cite{Nelson:2008uy} and Schwarzschild solutions, which
makes it difficult to place restrictions on these terms via cosmology
or solar-system tests. The best constraint on, different {\it ad hoc},
curvature squared terms are obtained from measurements of the orbital
precession of Mercury, imposing a rather weak lower bound on $\beta$,
namely $\beta > 3.2\times 10^{-9} {\rm m}^{-1}$~\cite{Stelle}.  This
constraint was however found for terms of different form (but of the
same order) to the Weyl term appearing in the NCG spectral action
approach we investigate here. In what follows, we will specifically
study how one can constrain $\beta$ within the NCG context. The
parameter $\beta$ can be equivalently expressed in terms of $f_0$,
through $\beta^2=(5\pi)/(48 G f_0)$, so by imposing a lower limit to
$\beta$, we actually set an upper limit to the moment $f_0$ of the
cut-off function used to define the spectral action. The normalization
of kinetic terms in the spectral action imposes the following relation
between the gauge couplings of the Standard Model, $g_1, g_2, g_3$ and
the coefficient $f_0$~\cite{ccm}, namely
$g_3^2f_0/( 2\pi^2)=1/ 4~,~g_3^2=g_2^2=(5/ 3)g_1^2$.
The importance of constraining $\beta$ is thus clear, since $f_0$ can
be used to specify the initial conditions on the gauge couplings, a
constraint on $\beta$ corresponds to a restriction on the particle
physics at unification.

We will study the energy lost to gravitational radiation by orbiting
binaries, so we consider the weak field limit of
Eq.~(\ref{eq:EoM2}). The general first order solution for a
perturbation against a Minkowski background is~\cite{NCG1}
\beq\label{eq:1stodrersol}
h^{\mu\nu}\left(\br,t\right) =\frac{4G\beta}{c^4} \int {\rm d}\br'
    {\rm d} t'\frac{\Theta\left(T\right)}{\sqrt{\left(cT\right)^2 - 
|\bR|^2}}\nonumber\\
 \times {\cal J}_1\left( \beta\sqrt{\left(cT\right)^2 - |\bR|^2}\right)
      T^{\mu\nu}\left( \br',t'\right)
\Theta\left(
cT - |\bR|\right) ~;
\eeq
$T=t-t'$ is the difference between the time of observation ($t$)
and emission ($t'$) of the perturbation, ${\bf R} = {\bf r} - {\bf
  r}'$ is the difference between the locations of the observer (${\bf
  r}$) and emitter (${\bf r}'$), ${\cal J}_1$ is a Bessel function of
the first kind and $\Theta$ is the Heavyside step function. In the far
field limit, $|\br| \approx |\br - \br'|$, the spatial components of
Eq.~(\ref{eq:1stodrersol}) become
\beq\label{eq:4} h^{ik}\left( \br,t\right) \approx \frac{2G
  \beta}{3c^4} \int_{-\infty}^{t-\frac{1}{c}|\br|} \frac{{\rm
    d}t'}{\sqrt{c^2\left( t-t'\right)^2 - |\br|^2} }\nonumber\\
\times {\cal J}_1 \left(
\beta\sqrt{c^2\left( t-t'\right)^2 - |\br|^2}\right)
\ddot{D}^{ik}\left(t'\right)~, \eeq
where we have, introduced the quadrupole moment,
\be
D^{ik}\left(t\right) \equiv \frac{3}{c^2}\int {\rm d}\br \ 
x^i x^k T^{00}(\br,t)~.
\ee
From Eq.~(\ref{eq:EoM2}) is it clear that this theory reduces to
that of GR in the $\beta\rightarrow \infty$ limit,
and one can check that in this limit Eq.~(\ref{eq:4}) does indeed
reproduce the standard result for a massless graviton. For finite
$\beta$ however, one finds that gravitational radiation contains both
massive and massless modes, both of which are sourced from the
quadrupole moment of the system. 


\section{Gravitational radiation from circular binaries}

We will derive the explicit formula for the energy lost to
gravitational radiation from a binary pair in a circular orbit.  One
can similarly consider binaries in elliptical orbits; for
simplicity we consider only circular Keplerian orbits. Similarly, we
neglect effects due to the internal structure of the bodies as well as
local astrophysical effects, such as mass transfer, tidal stripping
{\sl etc.}, focusing instead on the purely gravitational consequences of
NCG.

Consider a circular binary pair, of masses $m_1$, $m_2$. For such a
system, orbiting in the $xy$-plane, the only nonzero components of the
quadrupole moment are~\cite{landau_lifshitz}
\beq\label{eq:quad_mom}
 \ddot{D}^{xx}\left(t\right) 
  &=& 12 \mu |\rho|^2 \sin \left( 2\psi\left(t\right)\right)\omega^3
\nonumber\\
&=& -\ddot{D}^{yy}\left(t\right)~, \nonumber \\
 \ddot{D}^{xy}\left(t\right) &=& -12 \mu |\rho|^2 \cos\left( 2\psi\left(t\right)
\right)\omega^3~,\nonumber \\
D^{zz} &=& - \mu |\rho|^2~,
\eeq
%
where
$\mu = m_1m_2/(m_1+m_2)$ is the reduced mass of the system,
$|\rho|$ is the magnitude of the separation vector between the bodies,
which is constant for circular orbits, $\psi$ is the angle of the
bodies relative to the $x$-axis and $\omega=\dot{\psi}$ is the orbital
frequency, which for this simple system is a constant 
given by
\be \omega \equiv \dot{\psi} = |\rho|^{-3/2} \sqrt{ G\left( m_1 +
  m_2\right)}~.  \ee 
Following the standard approach ({\sl see e.g.},
Ref.~\cite{landau_lifshitz}), 
the rate of energy loss
, in the far field limit, is 
%
\be\label{eq:energy} -\frac{{\rm d} {\cal E}}{{\rm d}t} \approx
\frac{c^2}{20G} |{\bf r}|^2 \dot{h}_{ij} \dot{h}^{ij}~.  \ee
This allows us to explicitly test the theory by comparing this prediction
to binary pulsar measurements, for which the energy loss has been very well
characterized ({\sl see}, Table~\ref{tab:0}) and hence constrain $\beta$.

Using Eq.~(\ref{eq:4}) one finds~\cite{NCG1}
\beq
&&\dot{h}^{ij}\dot{h}_{ij}= \frac{128\mu^2|\rho|^4 \omega^6 G^2
  \beta^2}{c^8}\nonumber\\
&&~~~~~~~~ \times \left[ f_{\rm c}^2\left(\beta|{\bf
    r}|,\frac{2\omega}{\beta c}\right) + f_{\rm s}^2\left(\beta|{\bf
    r}|,\frac{2\omega}{\beta c}\right)\right]~, \eeq
 where we have
defined the functions:
\beq\label{eq:f1}
 f_{\rm s}\left( x,z\right) &\equiv& \int_0^\infty
\frac{{\rm d}s}{\sqrt{s^2 + x^2}} {\cal J}_1\left(s\right) \sin
\left(z\sqrt{ s^2 + x^2} \right)~,\nonumber\\
\label{eq:f2}
f_{\rm c}\left( x,z\right) &\equiv&
\int_0^\infty \frac{{\rm d}s}{\sqrt{s^2 + x^2}} {\cal
  J}_1\left(s\right) \cos \left(z\sqrt{ s^2 + x^2} \right).
\eeq
The integrals above, Eq.~(\ref{eq:f1}), exhibit a strong resonance
behavior at $z=1$, however they are easily evaluated for
both $z<1$ and $z>1$. This resonance corresponds to a critical
frequency given by 
\be
\label{critical}
2\omega_{\rm c} =\beta c~,
\ee
and we can expect strong deviations from the standard results of
GR for orbital frequencies close to this critical
frequency.

One can evaluate numerically the functions in Eq.~(\ref{eq:f1}) and
fit them to an explicit functional form. Thus, for $\omega<\omega_{\rm c}$ one
obtains
\beq\label{eq:fit} &&\Big[f_{\rm c}\left(\beta|{\bf
    r}|,\frac{\omega}{\omega_{\rm c}}\right)\Big]^2 + \Big[f_{\rm
    s}\left(\beta|{\bf r}|,\frac{\omega}{\omega_{\rm c}}\right)\Big]^2
\nonumber\\&&\approx \frac{1}{\left( \beta |{\bf r}|\right)^2} \exp\left(
\frac{C}{\beta |{\bf r}| \left( 1- \frac{\omega}{\omega_{\rm c}}\right) }
     {\cal J}_1\left( \beta |{\bf r}| -
     \frac{\omega}{\omega_{\rm c}}\right)\right)~, \nonumber \\ \eeq
where $C$ is approximately a constant, $C\approx 0.175$, except as
$\omega$ approaches $\omega_{\rm c}$. Similarly, for $\omega >
\omega_{\rm c}$ one gets
\beq &&\Big[f_{\rm c}\left(\beta|{\bf r}|,\frac{\omega}{\omega_{\rm
      c}}\right)\Big]^2 + \Big[f_{\rm s}\left(\beta|{\bf
    r}|,\frac{\omega}{\omega_{\rm c}}\right)\Big]^2 \\\approx &&
\frac{4}{\left( \beta |{\bf r}|\right)^2 } \sin^2\left( \beta |{\bf
  r}| \left( \tilde{f}\left(\frac{\omega}{\omega_{\rm
    c}}\right)\right)^{-1} \right)~, \eeq
where the function $\tilde{f}$ is approximately
\be \tilde{f}\left( \frac{\omega}{\omega_{\rm c}}\right) \approx
4\sqrt{ \left(\frac{\omega}{\omega_{\rm c}}\right)^2 - 1 } + 2\exp
\left( -\sqrt{ \left(\frac{\omega}{\omega_{\rm c}}\right)^2
  -1}\right)~.  \ee
Note that as we will see, the precise form of this function is
unimportant. See Ref.~\cite{NCG1} for a discussion on the accuracy of the
approximations.

Using the above approximations, one can expand Eq.~(\ref{eq:energy})
in the large distance (large $|{\bf r}|$) limit, to find the rate of
energy lost to gravitational radiation:
\beq\label{eq:GR_approx} -\frac{{\rm d} {\cal E}}{{\rm d}t} \approx
\frac{32 G \mu^2 \rho^4 \omega^6}{5c^5}
\ \ \ \ \ \ \ \ \ \ \ \ \ \ \ \ \ \ \ \ \ \ \ 
\ \ \ \ \ \ \ \ \ \ \ \ \ \ \ \ \ \nonumber
\\ \times\left\{ \begin{array}{cc} 1 + \frac{C}{\beta|{\bf r}| \left(
    1- \frac{\omega}{\omega_{\rm c}}\right)} {\cal J}_1 \left( \beta
  |{\bf r}| - \frac{\omega}{\omega_{\rm c}}\right) +\dots &
   ; \omega < \omega_{\rm c} \\ 4\sin^2\left( \beta|{\bf r}|
  \tilde{f}\left(\frac{\omega}{\omega_{\rm c}}\right)\right) & ;  \omega
  > \omega_{\rm c}
\end{array} \right.,
\eeq
where in the $\omega < \omega_{\rm c}$ case the dots refer to higher
powers of $1/\left(\beta |{\bf r}|\right)$. Thus, for orbital
frequencies small compared to $\omega_{\rm c}$, any deviation from the
standard result is suppressed by the distance to the source. Notice
that in this case, the $\beta \rightarrow \infty$ ({\sl i.e.},
$\alpha_0 \rightarrow 0$) limit reproduces the GR
result, as it should. For the $\omega > \omega_{\rm c}$ case, the
result would only agree with the General Relativistic result if $\beta
|{\bf r}|\tilde{f}\left(\omega/\omega_{\rm c}\right) = \pi/3$, which
is clearly not true for systems at different distances, $|{\bf r}|$,
with different orbital frequencies, $\omega$. Hence, we can
immediately eliminate the $\omega > \omega_{\rm c}$ case, simply by
noting that observations of the energy lost to gravitational radiation
agree, to a high level of accuracy, with those of GR
for many different systems.

The resonance appearing in Eq.~(\ref{eq:f1}) leads to a simple
heuristic argument to rule out the $\omega > \omega_{\rm c}$
case. A system with $\omega < \omega_{\rm c}$ cannot
increase its orbital frequency above $\omega_{\rm c}$, without losing
a significant fraction of its energy to gravitational radiation. Similarly,
a system with $\omega > \omega_{\rm c}$ cannot decrease its orbital frequency
across this boundary. Since one expects all astrophysical systems to have
formed from the coalescence of relatively cold, slowly moving systems,
it is reasonable to suppose that at some time in the past, all binary
systems had very slowly varying quadrupole moments and hence that
$\omega<\omega_{\rm c}$.  In the following, we will see that this
restriction places a strong constraint on $\beta$ (which defines
$\omega_{\rm c}$).

For the physically interesting case of $\omega<\omega_{\rm c}$, the
amplitude of the deviation from the standard result is small, due to
the $1/|{\bf r}|$ suppression, however there are two interesting
features: firstly, the existence of a critical frequency $\omega_{\rm
  c}$ and secondly, the fact that the rate of flux of gravitational
radiation is oscillatory.

The critical frequency comes from the fact that this theory contains a
natural frequency scale given by $\beta c \sim c\left( -\alpha_0
G\right)^{-1}$.  This scale is set by the moments of the cut-off
function used to define the spectral action. Physically one can think of this as the
scale at which noncommutative effects become dominant.  What is
important for this work, is that the binary systems must have orbital
frequencies below this critical value, since otherwise the theory
would predict significant deviations from the results of standard
GR, which can be ruled out observationally.

The presence of the Bessel function in Eq.~(\ref{eq:GR_approx}) means
that the amplitude of the deviation from the standard result of
GR will oscillate both with changing distances and
changing frequencies. Whilst such correlations may present new observational 
signatures, the effect is heavily suppressed by the $|{\bf r}|^{-1}$ factor
in Eq.~(\ref{eq:GR_approx}); here we focus on overall amplitude 
of deviations from the GR result.


\section{Astrophysical constraints}

Having calculated the general form of the gravitational radiation from
binary systems, within this NCG theory of gravity, we can now
constrain the main parameter of the theory, $\beta$, via observational
data. From Eq.~(\ref{eq:GR_approx}) it is clear that the only data
needed is the orbital frequency and the distance to the binary
system. If one had considered elliptical binary orbits, additional
parameters would come into play, however we are concerned only with
the order of magnitude of the constraint and hence neglect such
additional complications. A more detailed quantitative analysis would
require the inclusion of the ellipticity as well as various other near
field effects.

Table~\ref{tab:0} gives the binaries we are considering. We focus on
binary pulsars for which the rate of change of the orbital frequency
has been well characterized. In all these cases the predictions of
GR agree with the data to high accuracy ({\sl see},
Table~\ref{tab:0}).  We can thus restrict $\beta$ by requiring that
the magnitude of deviations from GR, given by
Eq.~(\ref{eq:GR_approx}), be less than this uncertainty.

\begin{table}[h]
\begin{center}
\begin{tabular}{|c|c|c|c|c|}
\hline
Binary & Distance & Orbital & Eccentricity & GR  \\
& (pc) & Period (hr) & & ($\%$)\\
\hline
PSR J0737-3039 & $\sim$ 500 & 2.454 & 0.088 & $ 0.2 $ \\
PSR J1012-5307 & $\sim$ 840 & 14.5 & $<10^{-6}$ & $ 10 $ \\
PSR J1141-6545 & $>$ 3700 & 4.74 & 0.17 & $ 6 $ \\
PSR B1916+16 & $\sim$ 6400 & 7.752 & 0.617 & $ 0.1 $ \\
PSR B1534+12 & $\sim$ 1100 & 10.1 & ? & $ 1 $ \\
PSR B2127+11C & $\sim$ 9980 & 8.045 & 0.68 & $ 3 $ \\
\hline
\end{tabular}
\caption{\label{tab:0} We calculate the constraint on the
  NCG theory of gravity, via the predicted energy
  lost to gravitational radiation from the above binaries (Refs.~\cite{Kramer:2006nb},~\cite{Lazaridis:2010hw},~\cite{Bhat:2008ck},~\cite{Taylor},~\cite{Stairs:1997kz},~\cite{Jacoby:2006dy},
  respectively). The column marked GR, indicates the approximate
  accuracy to which the rate of change of the orbital period agrees
  with the predictions of GR.}
\end{center}
\end{table}

Using the data on the six binaries given in Table~\ref{tab:0}, and
requiring that $\omega < \omega_{\rm c}$ ({\sl see}, the discussion
above), we find $\beta > 7.55\times 10^{-13}~{\rm m}^{-1}$. The
restrictions coming from the individual systems are given in
Table~\ref{tab:0.5}.
\begin{table}
\begin{tabular}{|c|c|}
\hline
PSR ~J0737-3039~ & ~~$\beta > 7.55\times 10^{-13}~{\rm m}^{-1}\ $ \\
PSR ~J1012-5307~ & ~~$\beta > 7.94\times 10^{-14}~{\rm m}^{-1}\ $ \\
PSR ~J1141-6545~ & ~~$\beta > 3.90\times 10^{-13}~{\rm m}^{-1}\ $ \\
PSR ~~B1913+16~   & ~~$\beta > 2.39\times 10^{-13}~{\rm m}^{-1}\ $ \\
PSR ~~B1534+12~   & ~~$\beta > 1.83\times 10^{-13}~{\rm m}^{-1}\ $ \\
\ PSR ~B2127+11C~  & ~~$\beta > 2.30\times 10^{-13}~{\rm m}^{-1}\ $ \\
\hline
\end{tabular}
\caption{\label{tab:0.5} For each binary system, we restrict $\beta$
by requiring that the energy lost to gravitational radiation agrees
with the prediction of GR to within observational
uncertainties.}
\end{table}

Due to the large distances to these systems, the constraint is almost
exactly due to $\omega < \omega_{\rm c}$ which, using the definition
of $\omega_{\rm c}$ given in Eq.~(\ref{critical}), becomes $\beta >
2\omega /c$. Thus, the strongest constraint comes from systems with
high orbital frequencies.  This will be true for all systems for which
$2\omega |{\bf r}|/c$ is large.  Future observations of rapidly
orbiting binaries, relatively close to the Earth, could thus improve
this constraint by many orders of magnitude.

This dependence of the constraint on the orbital frequency, suggests
that other astrophysical objects, with high frequency periodicity,
such as individual pulsars or merger in-spirals may provide a more
stringent constraint. Whilst the analysis given here is only
applicable to binaries, it can be extended  by replacing
Eq.~(\ref{eq:quad_mom}) by the quadrupole moments of whatever system
is of interest.

\section{Conclusions}
General Relativity is formulated within the arena of Riemannian
geometry a natural extension of which is NonCommutative Geometry. The
spectral action approach produces all the Standard Model fields as
well as gravitational terms, from purely geometric considerations.
Thus, both gravity and matter are treated in a similar manner within
NCG, which also provides us with concrete relationships between matter
and gravitational couplings. The asymptotic expansion of the
gravitational sector of this theory produced modifications to GR
and in this paper we use these modifications to test and
constrain the theory through observations.

We have considered the energy lost by circular binary systems to
gravitational radiation and shown that for the predicted values to
agree with observations, a key parameter of the theory can be
constrained. We have focused on 
binary pulsar systems, for
which the rate of change of the orbital frequency is well known and
explicitly calculated the predicted deviation from the GR
expressions. We have shown that this restricts the value of the Weyl
squared coupling in the bosonic action ({\sl i.e.}, $\alpha_0$ in
Eq.~(\ref{eq:0})). This observational constraint may seem rather weak,
requiring only that $\beta \geq 7.55\times 10^{-13} {\rm m}^{-1}$,
however it is comparable to (but larger than) existing constraints on
similar, {\sl ad hoc}, additions to GR. In particular, constraints on
additions to the Einstein-Hilbert action, of the form $R^2$ and
$R_{\mu\nu} R^{\mu\nu}$, are of the order of $\beta_{R^2} \geq
3.2\times 10^{-9} {\rm m}^{-1}$, where $\beta_{R^2}$ is the $\beta$
parameter associated with the couplings of these
terms~\cite{Stelle}. Whilst our constraint is several orders of
magnitude weaker than these, it will rapidly be improved as more
binary pulsars are discovered and the observations of existing systems
improve. This is to be contrasted with the existing constraints which
rely on the perihelion precession of Mercury, the accuracy of which is
unlikely to improve significantly in the future.

As an example, white dwarf binaries reach orbital frequencies of the
order of $\sim (10-100){\rm mHz}$ towards the end of their merger,
whilst neutron binaries reach $\sim (10-100){\rm Hz}$ and should be
readily observable with the {\sl Laser Interferometer Space Antenna}
(LISA)~\cite{Finn}. Whilst such systems would require a greater
understanding of the strong and near field effects than that presented
here, one can expect the constraint on $\beta$ coming from such
objects to be of the order of $\beta > (10^{-10}-10^{-6}){\rm
  m}^{-1}$.

We were able to constrain the natural length, defined through
the $f_0=f(0)$ momentum of the cut-off function $f$ --- a real parameter
related to the coupling constants at unification --- at which the
noncommutative effects become dominant, by purely astrophysical
observations.

\section{Acknowledgments}
The work of W.\ N.\ is partially supported  by the NSF grant PHY0854743,
the G.\ A. \& M.\ M. Downsbrough Endowment and the Eberly
research fund of Penn State.  The work of M.\ S.\ is partially
supported by the European Union through the Marie Curie Research \&
Training Network {\sl UniverseNet} (MRTN-CT-2006-035863). J.\ O.\ 
acknowledges support from the Alfred P. Sloan Foundation and the 
Eberly College of Science.


\end{document}